\documentstyle[aps,prb]{revtex}
\begin{document}
\draft
\title{Flux creep and flux jumping}
\author{R.G.~Mints}
\address{School of Physics and Astronomy,\\ Raymond
and Beverly Sacler Faculty of Exact Sciences,\\ Tel Aviv
University, Tel Aviv 69978, Israel}
\maketitle
\begin{abstract}
We consider the flux jump instability of the Bean's 
critical state arising in the flux creep regime in type-II
superconductors. We find the flux jump field, $B_j$, that
determines the superconducting state stability criterion. 
We calculate the dependence of $B_j$ on the external 
magnetic field ramp rate, $\dot B_e$. We demonstrate that 
under the conditions typical for most of the magnetization 
experiments the slope of the current-voltage curve in the 
flux creep regime determines the stability of the Bean's 
critical state, {\it i.e.}, the value of $B_j$. We show 
that a flux jump can be preceded by the magneto-thermal 
oscillations and find the frequency of these oscillations
as a function of $\dot B_e$.
\end{abstract}
\pacs{74.60. Ec, 74.60. Ge}

\narrowtext
\section{Introduction}
The Bean's critical state model\cite{bea} successfully 
describes the irreversible magnetization in type-II 
superconductors by introducing the critical current density
$j_c(T,B)$, where $T$ is the temperature and $B$ is the 
magnetic field. In the framework of the Bean's model the value 
of the slope of the stationary magnetic field profile is 
less or equal to $\mu_0j_c(T,B)$. This nonuniform flux 
distribution does not correspond to an equilibrium state and 
under certain conditions flux jumps arise in the critical 
state. The flux jumping process results in a flux 
redistribution towards the equilibrium state and is 
accompanied by a strong heating of the superconductor.
\par
Flux jumping has been numerously studied in conventional 
and high-temperature superconductors (see the review papers
\cite{min_1,wip}, references therein, and the recent
experimental studies \cite{hen,leg,ger,lei}). In the general 
case two types of flux jumps can be considered, namely, the 
global and the local flux jumps. A global flux jump involves 
vortices into motion in the entire volume of the sample. A 
local flux jump happens in a small fraction of the sample 
volume. Depending on the initial perturbation and the driving 
parameters there are two qualitatively different types of 
global flux jumps, namely, complete and partial flux jumps. 
The first turns the superconductor to the normal state. The 
second self-terminates when the temperature is still less 
than the critical temperature. 
\par
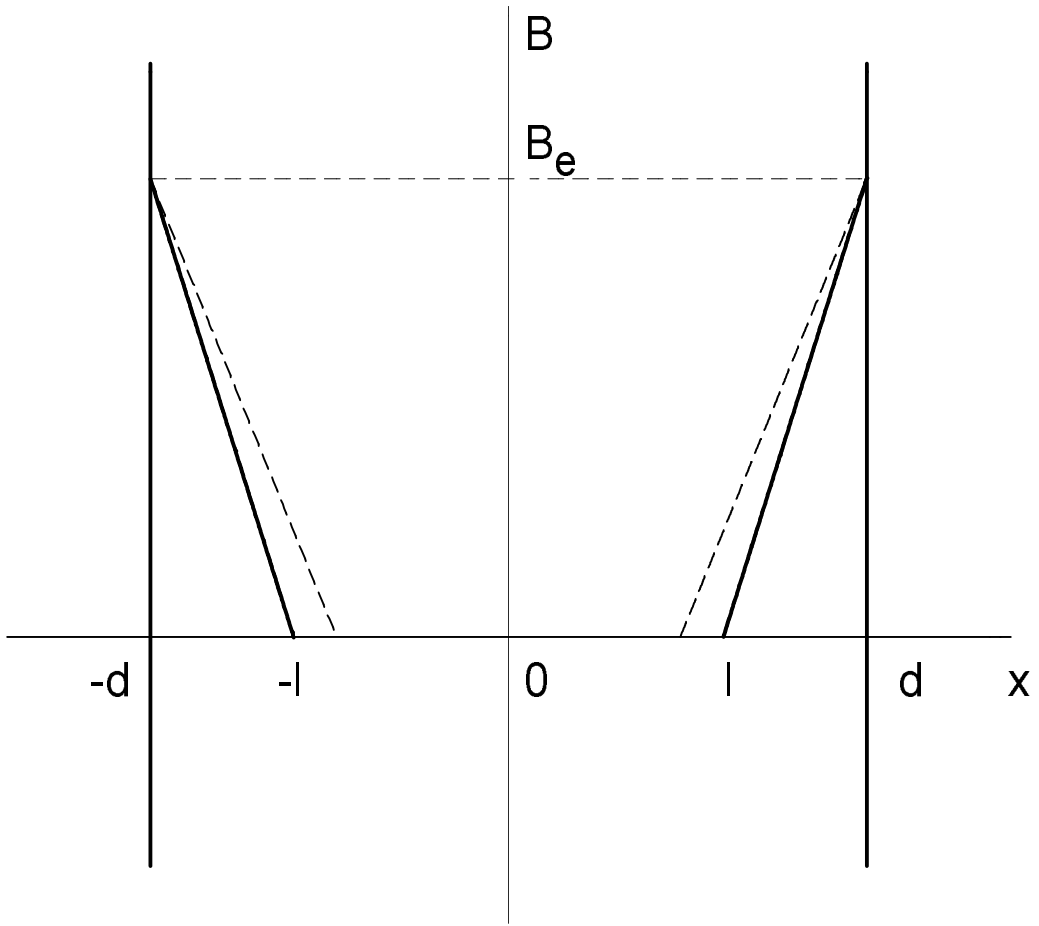
\begin{figure}
\caption{Magnetic field $B(x)$ distribution at different 
temperatures: $T=T_0$ (solid line), $T=T_0+\delta T$ (dashed 
line).}
\label{f1}
\end{figure}

We illustrate a global flux jump origination treating a 
superconducting slab with the thickness $2d$ subjected to an 
external magnetic field ${\bf B}_e$ parallel to the sample 
surface ($yz$ plane). In the framework of the Bean's 
critical state model the space distribution of the flux is 
given by the equation 
\begin{equation}
{dB\over dx}=\pm\,\mu_0 j_c,
\label{e1} 
\end{equation}
where the $\pm$ stays for $x>0$ and $x<0$ correspondingly.
We show the dependence $B(x)$ in Fig.~\ref{f1} for the case, 
when the critical current density depends only on the
temperature, {\it i.e.}, $j_c=j_c(T)$. 
\par
Let us now suppose that the temperature of the sample $T_0$ 
is increased by a small perturbation $\delta T_0$
arising due to a certain initial heat release $\delta Q_0$. 
The critical current density $j_c(T)$ is a decreasing 
function of temperature. Thus, the density of the 
superconducting current screening of the external magnetic 
field at $T=T_0+\delta T_0$ is less than at $T=T_0$. This 
reduction of the screening current leads to a rise of the 
magnetic flux inside the superconductor as it is shown in 
Fig.~\ref{f1}. The motion of the magnetic flux into the 
sample, which occurs as a result of the temperature 
perturbation $\delta T_0$, induces an electric field 
perturbation $\delta\! E_0$. The arise of $\delta\! E_0$ is 
accompanied by an additional heat release $\delta Q_1$, an 
additional temperature rise $\delta T_1$, and, therefore, an 
additional reduction of the superconducting screening 
current density $j_c$. Under certain conditions it results 
in an avalanche-type increase of the temperature and the 
magnetic flux in the superconductor, {\it i.e.}, in a 
global flux jump.
\par
The relative effect of the flux and temperature 
redistribution dynamics on the flux jumping process depends 
on the ratio, $\tau$, of the flux, $t_m$, and thermal, 
$t_{\kappa}$, diffusion time constants \cite{min_1}. The 
value of the dimensionless parameter $\tau$ is determined by 
the corresponding diffusion coefficients and is equal to
\begin{equation}
\tau=\mu_0\,{\lambda\sigma\over C},
\label{e2}
\end{equation}
where $\lambda$ is the heat conductivity, $\sigma$ is the 
conductivity, and $C$ is the heat capacity. 
\par
For $\tau\ll 1$ ($t_m\ll t_{\kappa}$), rapid propagation 
of the flux is accompanied by an adiabatic heating of the 
superconductor, {\it i.e.}, there is not enough time to
redistribute and remove the heat released due to the flux 
motion. For $\tau\gg 1$ ($t_{\kappa}\ll t_m$), the space 
distribution of flux remains fixed during the stage of 
rapid heating. These adiabatic ($\tau\ll 1$) and dynamic 
($\tau\gg 1$) approximations are the basis of the approach 
to the flux jumping problem \cite{min_1} and a flux jump 
scenario significantly depends on the relation between the
values of the heat conductivity $\kappa$, heat capacity 
$C$, and conductivity $\sigma$, that is determined by 
the slope of the $j$-$E$ curve.
\par
Let us now estimate the electric field value typical for 
the magnetization experiments. In this case the external 
magnetic field ramp rate $\dot B_e$ is usually from the 
interval $\dot B_e<1$~Ts$^{-1}$. The background 
electric field, $E_b$, induced by the magnetic field 
variation is of the order of $E_b\sim\dot B_e(d-l)$, 
where $d-l$ is the width of the area occupied by the 
critical state (see, for example, Fig.~\ref{f1}). We 
estimate $E_b$ as $E_b<10^{-6}$~Vcm$^{-1}$ using the value 
$d-l<10^{-4}$~m which is typical for the stability domain 
of the Bean's critical state. This electric field interval 
corresponds to the flux creep regime. Therefore, for the 
magnetization experiments the background electric field 
$E_b$ is from the flux creep regime, where the relation 
between the current density, $j$, and the electric field, 
$E$, is strongly nonlinear. As a result, the value of 
$\sigma$, {\it i.e.}, the slope of the $j$-$E$ curve, is 
strongly electric field dependent and the flux jumping 
takes place on a background of a resistive state with a 
conductivity that strongly depends on the external magnetic 
field ramp rate $\dot B_e$. 
\par
In order to calculate the conductivity in the flux creep 
regime we use the dependence of $j$ on $E$ in the form
\begin{equation}
j=j_c+j_1\ln\Bigl({E\over E_0}\Bigr),
\label{e3}
\end{equation}
where $E_0$ is the voltage criterion at which the critical 
current density $j_c$ is defined, $j_1$ determines the 
slope of the $j$-$E$ curve and $j_1\ll j_c$. Note, that the actual 
choice of $E_0$ is not very essential. Indeed, by taking for 
the voltage criterion a certain value $\tilde E_0$ instead of 
$E_0$ we change the critical current density from $j_c$ to 
$\tilde j_c=j_c-j_1\ln(\tilde E_0/E_0)$. The difference 
between $\tilde j_c$ and $j_c$ is small as 
$\ln(\tilde E_0/E_0)\sim 1$ and $j_1\ll j_c$. It is common to 
define the critical current value as the current density at 
$E_0=10^{-6}$~Vcm$^{-1}$. 
\par
Let us also note, that a power law
\begin{equation}
j=j_c\,\Bigl({E\over E_0}\Bigr)^{1/n}
\label{e3a} 
\end{equation}
with $n\gg 1$ is often used to describe the $j$-$E$ curve 
in the flux creep regime. Expanding the dependence given 
by Eq.~(\ref{e3a}) in series in $1/n\ll 1$ and keeping 
the first two terms we find that if we take $n=j_c/j_1$,
then Eqs.~(\ref{e3}) and (\ref{e3a}) coincide with the 
accuracy of $1/n^2\ll 1$.
\par
The relation given by Eq.~(\ref{e3}) was first derived in 
the framework of the Anderson-Kim model 
\cite{and_1,kim,and_2} considering the thermally activated 
uncorrelated hopping of bundles of vortices. The 
vortex-glass \cite{fis} and collective creep \cite{fei,nat} 
models result in more sophisticated dependencies of $j$ on 
$E$. However, these $j$-$E$ curves coincide with the one 
given by Eq.~(\ref{e3}) if $j-j_c\ll j_c$. The recently 
developed self-organized criticality approach to the 
critical state \cite{wan,pan} also results in 
Eq.~(\ref{e3}) if $j-j_c\ll j_c$. The logarithmic 
dependence of the current density $j$ on the electric 
field $E$ in the interval $j-j_c\ll j_c$ is in a good 
agreement with numerous experimental data \cite{gur}. In 
this paper we use the $j$-$E$ curve given by 
Eq.~(\ref{e3}) to calculate the conductivity $\sigma$ 
assuming that $j_1/j_c\ll 1$.
\par
It follows from Eq.~(\ref{e3}) that for the flux creep 
regime the conductivity $\sigma$ is given by the formula
\begin{equation}
\sigma=\sigma (E)={dj\over dE}={j_1\over E}.
\label{e4} 
\end{equation}
We estimate the value of $\sigma$ as 
$\sigma>10^{10}$~$\Omega^{-1}$cm$^{-1}$ using the typical 
data $j_1>10^{3}$~Acm$^{-2}$ and $E<10^{-7}$~Vcm$^{-1}$.
It follows from this estimation that the conductivity 
$\sigma$ determining the flux jumps dynamics for the 
magnetization experiments is very high. As a consequence 
the dimensionless ratio $\tau$ is also very high. Thus, the 
scenario of a flux jump for the magnetization experiments 
corresponds to the limiting case when $\tau\gg 1$ and the 
rapid heating stage takes place on the background of a 
``frozen-in'' magnetic flux.
\par
The electric field dependent conductivity $\sigma (E)$ 
significantly affects the flux jumping process. In 
particular, it results in the flux jump field $B_j$ 
dependence on the magnetic field ramp rate $\dot B_e$. This 
dependence is known from experiments \cite{min_1,wip,leg} 
and was never considered theoretically as a consequence of 
the logarithmic $j$-$E$ curve characterizing the flux creep 
regime in superconductors with high values of the critical 
current density $j_c$.
\par
Under certain conditions a flux jump is preceded by a 
series of magneto-thermal oscillations \cite{min_1}. These 
oscillations have been observed for the low-temperature 
\cite{zeb,chi} as well as for the high-temperature 
superconductors \cite{leg}. Theoretically magneto-thermal 
oscillations were considered for a flux jump developing in 
the flux flow regime \cite{min_3}. In this case the $j$-$E$ 
curve is linear and the value of the conductivity $\sigma$ 
is electric field independent. The high and electric field 
dependent conductivity $\sigma (E)$ significantly affects
the flux dynamics and therefore the magneto-thermal 
oscillations. In particular, it results in the dependence 
of the frequency of the magneto-thermal oscillations on the 
magnetic field ramp rate $\dot B_e$. The effect of the 
logarithmic $j$-$E$ curve on the magneto-thermal 
oscillations was never treated theoretically.
\par
In this paper we consider the flux jump instability of 
the Bean's critical state arising in the flux creep 
regime on the background of a nonuniform electric field
determining the conductivity of the type-II superconductor
in the flux creep regime. We find the flux jump field, 
$B_j$, that determines the critical state stability 
criterion and the dependence of $B_j$ on the external 
magnetic field, $B_e$, and the external magnetic field 
ramp rate, $\dot B_e$. We show that a flux jump can be 
preceded by the magneto-thermal oscillations and find 
the frequency of these oscillations as a function of 
$\dot B_e$.
\par
The paper is organized in the following way. In Sec.~II, 
we consider the critical state stability qualitatively
and obtain the stability criterion. In Sec.~III, we 
derive the equations determining the development of the 
small temperature and electric field perturbations and
calculate the frequency of the magneto-thermal 
oscillations. In Sec.~IV, we summarize the overall 
conclusions.
\par
\section{Qualitative consideration}
In this section we consider the critical state stability
qualitatively assuming that the thermomagnetic instability
develops much faster than the magnetic flux diffusion
process. In other words, we treat the case when the heating 
accompanying the thermomagnetic instability takes place on 
the background of a ``frozen-in'' magnetic flux. In 
Sec.~III we derive the exact criterion of applicability 
of the following qualitative reasoning.
\par
Let us consider a superconducting slab with the thickness 
$2d$ subjected to a magnetic field parallel to the sample 
surface (see Fig.~\ref{f1}) and suppose that the temperature 
of the sample $T_0$ is increased by a small perturbation 
$\delta T$. To keep the critical state stable, {\it i.e.}, 
to keep the screening current at the same level, an electric 
field perturbation $\delta\! E$ arises. The additional 
electric field $\delta\! E$ causes an additional heat 
release $\delta Q\propto\delta\! E$, which is the ``price'' 
for keeping the total screening current density at the same 
level, {\it i.e.}, the ``price'' for the ``frozen-in'' 
magnetic flux. 
\par
The critical state is stable if the additional heat 
release $\delta Q$ can be removed to the coolant by the 
additional heat flux $\delta W\propto\delta T$ resulting 
from the temperature perturbation $\delta T$. Thus, the 
critical state stability criterion has the form
\begin{equation} 
\delta W > \delta Q.
\label{e5} 
\end{equation}
\par
The additional heat release per unit length, $\delta Q$,
is given by the integral of $j\delta\! E$
over the width of the superconducting slab
\begin{equation} 
\delta Q=\int\limits_{-d}^{d} 
j\,\delta\! E\,dx.
\label{e6} 
\end{equation}
\par
The additional heat flux, $\delta W$, is determined by the 
temperature perturbation $\delta T$ at the sample surface,
{\it i.e.}, 
\begin{equation} 
\delta W=h\,\delta T\Big\vert_P,
\label{e7} 
\end{equation}
where $h$ is the heat transfer coefficient to the coolant 
with the temperature $T_0$ and $P$ stays for the sample 
surface.
\par
Using Eqs.~(\ref{e5}), (\ref{e6}) and (\ref{e7}) we find 
the critical state stability criterion, namely, the
inequality
\begin{equation} 
\int\limits_{-d}^{d} j\,\delta\! E\,dx < 
2h\,\delta T\Big\vert_P. 
\label{e8} 
\end{equation}
\par
To derive the explicit form of this stability criterion we
have to find the relation between $\delta T$ and 
$\delta\! E$. To do it, we calculate the decrease of the 
current density $\delta j_-$ resulting from the temperature 
perturbation $\delta T$ and the increase of the current 
density $\delta j_+$ resulting from the electric field 
perturbation $\delta\! E$. If the critical state is
stable then the total screening current density stays 
constant. As a result, the relation between $\delta\! E$ and
$\delta T$ is given by the equation
\begin{equation} 
\delta j=\delta j_-+\delta j_+=0.
\label{e9} 
\end{equation}
\par
In the critical state, $j\approx j_c$, thus, the 
decrease of $j$ due to the temperature perturbation 
$\delta T$ is equal to
\begin{equation}
\delta j_-=
-\Bigl\vert{\partial j_c\over\partial T}\Bigr\vert\,
\delta T
 \label{e10} 
\end{equation}
(note that ${\partial j_c/\partial T}<0$).
\par
The increase of the current density due to the electric
field perturbation $\delta\! E$ can be written as
\begin{equation} 
\delta j_+={dj\over dE}\,\delta\! E=
\sigma\,\delta\! E.
\label{e11} 
\end{equation}
Note, that the conductivity $\sigma$ is the differential
conductivity, {\it i.e.}, it is determined by the slope
of the $j$-$E$ curve.
\par
Combining Eqs.~(\ref{e4}) and (\ref{e11}), we find the 
relation between $\delta j_+$ and $\delta\! E$ in the form
\begin{equation} 
\delta j_+={j_1\over E_b}\,\delta\! E=
{j_c\over nE_b}\,\delta\! E,
\label{e12} 
\end{equation}
where $n=j_c/j_1\gg 1$. 
\par
It follows from Eqs.~(\ref{e9}), (\ref{e10}) and 
(\ref{e12}) that
\begin{equation} 
\delta\! E={1\over\sigma}\,
\Bigl\vert{\partial j_c\over\partial T}\Bigr\vert\,
\delta T={nE_b\over j_c}\,
\Bigl\vert{\partial j_c\over\partial T}\Bigr\vert\,
\delta T.
\label{e13} 
\end{equation}
\par
Equations (\ref{e4}) and (\ref{e13}) allow to understand 
the effect of the background electric field $E_b$ on the 
critical state stability. It follows from Eq.~(\ref{e4}) 
that a low electric field $E_b$ results in a high 
differential conductivity ($\sigma\propto 1/E_b$). 
In its turn a high conductivity $\sigma$ leads to a low 
electric field perturbation (indeed, it follows from 
Eq.~(\ref{e13}) that 
$\delta\! E\propto 1/\sigma\propto E_b$). The smaller is 
$\delta\! E$, the less ``costly'' it is to remove the 
additional heat release. As a result the lower is the 
background electric field $E_b$ the more stable is the 
superconducting state.
\par
Substituting Eq.~(\ref{e13}) into Eq.~(\ref{e8}) we find 
the critical state stability criterion in the form
\begin{equation} 
\int\limits_{-d}^{d} nE_b\,
\Bigl\vert{\partial j_c\over\partial T}
\Bigr\vert\,\delta T\, dx <
2h\,\delta T\Big\vert_P.
\label{e14} 
\end{equation}
\par
We have to treat the temperature perturbation 
$\delta T$ in more detail to derive the final form of 
Eq.~(\ref{e14}). The variation of the function 
$\delta T(x)$ on the interval $-d\le x\le d$ depends on the 
value of the Biot number
\begin{equation} 
Bi={dh\over\kappa},
\label{e15} 
\end{equation}
where $\kappa$ is the heat conductivity of the 
superconductor. Let us assume that the value of the 
heat transfer coefficient $h$ is relatively low. As a 
result, $Bi\ll 1$ and the temperature perturbation 
$\delta T(x)$ is almost uniform over the width of the 
superconducting slab. It means that $\delta T$ cancels in 
both sides of Eq.~(\ref{e14}) and the Bean's critical state 
stability criterion takes the following final form 
\begin{equation}
{\cal J}={n\over 2h}\,\int\limits_{-d}^{d} E_b\,
\Bigl\vert{\partial j_c\over\partial T}
\Bigr\vert\, dx < 1. 
\label{e16} 
\end{equation}
Let us note, that this criterion was first derived in order 
to calculate the maximum value of a superconducting current 
under conditions typical for the critical current 
measurements, {\it i.e.}, for a superconducting wire 
carrying a current that is increased with a given ramp 
rate \cite{min_2}. 

\par
In addition, we assume for simplicity that the value of $n$ 
is temperature and magnetic field independent if $T<T_c$ 
and $B<B_{c2}$, where $T_c$ is the critical temperature and
$B_{c2}$ is the upper critical field. This assumption is 
in a good agreement with numerous experimental data 
\cite{gur} as well as with the self-organized criticality 
approach to the Bean's critical state \cite{wan,pan}.
\par 
Using Eq.~(\ref{e1}) we can rewrite the criterion given by 
Eq.~(\ref{e16}) in the following form which is convenient 
for the further analysis
\begin{equation}
{\cal J}={n\over h}\,\int\limits_0^{d} E_b\,
\Bigl\vert{\partial j_c\over\partial T}
\Bigr\vert\, dx={n\over\mu_0 h}\,\int\limits_{B^*}^{B_e}
{E_b\over j_c}\,
\Bigl\vert{\partial j_c\over\partial T}
\Bigr\vert\, dB<1,
\label{e17} 
\end{equation}
where $B^*=B(0)$ is the magnetic field in the middle plane
of the superconducting slab.
\par 
The background electric field $E_b$ is induced by the 
varying external magnetic field $B_e(t)$ and thus the 
spatial distribution of $E_b$ is given by the Maxwell 
equation
\begin{equation} 
{dE_b\over dx}={dB\over dt}.
\label{e18} 
\end{equation}
Combining Eqs.~(\ref{e1}) and (\ref{e18}) and taking into 
account that $j\approx j_c$ we find that
\begin{equation}
{dE_b\over dB}=\pm\,{\dot B\over\mu_0 j_c(B)}, 
\label{e19}
\end{equation} 
where the $\pm$ stays for $x>0$ and $x<0$ correspondingly.
At the same time Eq.~(\ref{e18}) results in the relation
\begin{equation}
{\dot B_e\over j_c(B_e)}=
{\dot B\over j_c(B)}. 
\label{e20} 
\end{equation}
\par
It follows from Eqs.~(\ref{e19}) and (\ref{e20}) that the 
the background electric field $E_b$ dependence on $B$
is given by the formula 
\begin{equation}
E_b=\pm\,{\dot B_e(B-B^*)\over\mu_0 j_c(B_e)},
\label{e21} 
\end{equation}
where the $\pm$ stays for $x>0$ and $x<0$ correspondingly.
\par
Let us now apply the criterion given by Eq.~(\ref{e17}) 
to calculate the flux jump field $B_j$ assuming that 
initially there is no flux inside the superconducting slab, 
{\it i.e.}, we calculate now the magnetic field of the 
first flux jump. Using Eqs.~(\ref{e17}) and (\ref{e21}) we 
find the stability criterion in the form
\begin{equation}
{\cal J}={n\dot B_e\over\mu_0^2 hj_c(B_e)}\,
\int\limits_{B^*}^{B_e}{B-B^*\over j_c(B)}\,
\Bigl\vert{\partial j_c\over\partial T}\Bigr\vert\, dB<1.
\label{e22} 
\end{equation}
\par
The value of the magnetic field $B^*$ is given by the 
following system of equations
\begin{eqnarray}
B^*=0,\qquad &{\rm if}\quad B_e<B_p,\\
\label{e23} 
\int\limits_{B^*}^{B_e}{dB\over j_c(B)}=\mu_0 d,
\qquad &{\rm if}\quad B_e>B_p,
\label{e24} 
\end{eqnarray}
where the value of the magnetic flux penetration field, 
$B_p$, is determined by
\begin{equation} 
\int\limits_0^{B_p}{dB\over j_c(B)}=\mu_0d.
\label{e25} 
\end{equation}
\par
It follows from Eqs.~(\ref{e22}) and (\ref{e24}) that 
$\cal J$ is an increasing function of the external 
magnetic field $B_e$ if $B_e<B_p$ and $\cal J$ is a 
decreasing function of $B_e$ if $B_e>B_p$. In other words,
if for a given value of $\dot B_e$ the superconducting 
state is stable in the region $0<B_e<B_p$ then it is stable 
for any magnetic field. Thus, if a flux jump occurs it 
occurs only if $B_e<B_p$. Therefore, we consider now a
superconducting slab that is wide enough meaning that 
$B_j<B_p$.
\par   
We have the criterion ${\cal J}(B_j)=1$ to find the flux 
jump field $B_j$ in the case when $B_e<B_p$. Thus, it 
follows from Eq.~(\ref{e22}) that the dependence 
$B_j(\dot B_e)$ is given by the equation
\begin{equation}
{\cal J}(B_j)={n\dot B_e\over\mu_0^2 hj_c(B_j)}\,
\int\limits_0^{B_j}{B\over j_c(B)}\,
\Bigl\vert{\partial j_c\over\partial T}\Bigr\vert\, dB=1. 
\label{e26} 
\end{equation}
\par
Let us approximate the value of 
$\vert\partial j_c/\partial T\vert$ as
\begin{equation}
\Bigl\vert{\partial j_c\over\partial T}\Bigr\vert \approx
{j_c(B)\over T_c(B)-T_0}.
\label{e27} 
\end{equation}
Using Eq.~(\ref{e27}) we rewrite Eq.~(\ref{e26}) in the 
following form 
\begin{equation}
{n\dot B_e\over\mu_0^2 hj_c(B_j)}\,\int\limits_0^{B_j}
{B\over T_c(B)-T_0}\, dB=1.
\label{e28} 
\end{equation}
\par
We treat now the case when $T_0\ll T_c(B_j)$ or in other 
words $B_j\ll B_{c2}(T_0)$. It means that
$T_c(B)\approx T_c$, where $T_c$ is the critical 
temperature at zero magnetic field. It follows finally 
from Eq.~(\ref{e28}) that the Bean's critical state 
stability criterion determining the the dependence 
$B_j(\dot B_e)$ is given by the equation  
\begin{equation}
{B_j^2\over j_c(B_j)}=
{2\mu_0^2\,h(T_c-T_0)\over n\dot B_e}.
\label{e29} 
\end{equation}
\par
Let us now consider the particular case of the Bean's 
critical state model, namely, let us assume that the 
critical current density is magnetic field independent,
{\it i.e.},
\begin{equation}
j_c=j_c(T).
\label{e30} 
\end{equation}
\par
Using Eq.~(\ref{e29}) we find the following formula for 
the first flux jump field $B_j$
\begin{equation}
B_j=\sqrt{2\mu_0^2\,j_c(T_0)\,h(T_c-T_0)\over n\dot B_e}
\propto {1\over\dot B_e^{\,1/2}}.
\label{e31}
\end{equation}
\par
It follows from Eq.~(\ref{e31}) that the value of $B_j$ 
is inversely proportional to the square root of the 
magnetic field ramp rate $\dot B_e$ and is, therefore, 
decreasing with the increase of $\dot B_e$. The physics of 
this effect is related to the decrease of the conductivity 
$\sigma (E)$ in the flux creep regime with the increase of 
the background electric field $E_b$, {\it i.e.}, with the 
increase of $\dot B_e$. 
\par
We derive the expression for $B_j$ assuming that the rapid 
heating stage of a flux jump takes place on the background 
of a ``frozen-in'' magnetic flux. This approach is valid if
$\tau\gg 1$ which is the same as
\begin{equation}
\dot B_e\ll{1\over n}\,{B_p\over t_\kappa},
\label{e35} 
\end{equation}
where we introduce the typical thermal diffusion time 
constant $t_\kappa$ as
\begin{equation}
t_\kappa={d^2C\over\kappa}.
\label{e34} 
\end{equation}
\par
Let us now compare the values of $B_j$ and $B_a$, where 
$B_a$ determines the flux jump field for the adiabatic 
stability criterion \cite{han}. This well known criterion 
is based on the suggestion that the heating accompanying a 
flux jump is an adiabatic process, {\it i.e.}, it is 
assumed that there is no heat redistribution during a flux 
jump. Therefore, the adiabatic stability criterion 
corresponds to the limiting case of $\tau\ll 1$ that is
not the case typical for the magnetization experiments.
\par
The value of $B_a$ is given by the formula \cite{han,min_1}
\begin{equation}
B_a={\pi\over 2}\,\sqrt{\mu_0C(T_0)(T_c-T_0)}.
\label{e32} 
\end{equation}
It follows from the comparison of Eqs.~(\ref{e31}) and 
(\ref{e32}) that $B_a<B_j$ if  
\begin{equation}
\dot B_e<{8\over\pi^2}\,{\mu_0j_1h\over C}=
{8\over\pi^2}\,{Bi\over n}\,{B_p\over t_\kappa}.
\label{e33} 
\end{equation}
Note, that the inequality given by Eq.~(\ref{e33}) is 
stronger than the one given by Eq.~(\ref{e35}) as we 
assume that $Bi\ll 1$.
\par
The critical current density is decreasing with the 
increase of the magnetic field. Let us now consider the 
effect of this dependence on the critical state stability 
assuming that the value of $B_j$ is relatively high. To 
do it we use the Kim-Anderson model \cite{and_2} to 
describe the function $j_c(B)$. In the case of a high 
magnetic field this model results in the relation
\begin{equation}
j_c={\alpha (T)\over B}.
\label{e36}
\end{equation}
\par
Using Eqs.~(\ref{e36}) and (\ref{e29}) we find for the 
flux jump field $B_j$ the formula
\begin{equation}
B_j=\Biggl({2\mu_0^2\alpha (T_0)h(T_c-T_0)
\over n\dot B_e}\Biggr)^{1/3}
\propto{1\over\dot B_e^{\,1/3}}.
\label{e37}
\end{equation}
\par
The comparison of Eqs.~(\ref{e37}) and (\ref{e31}) 
shows that the magnetic field dependence of the 
critical current density slows down the decrease of $B_j$
with the increase of $\dot B_e$.

\section{Quantitative consideration} 
We treat now the critical state stability  in more detail 
and, in particular, we take into consideration the 
magneto-thermal oscillations. We consider a superconducting 
slab with the thickness $2d$ subjected to an external 
magnetic field parallel to the $z$ axis (see Fig.~\ref{f1}). 
\par
We use for calculation the Bean's critical state model 
assuming that the critical current density is magnetic 
field independent, {\it i.e.}, $j_c=j_c(T)$. We suppose 
also that $B_e\le B_p=\mu_0j_cd$. The  space distribution 
of the background electric field $E_b$ is then given by 
the formulae
\begin{equation}
E_b(x)=\cases{
\dot B_e(x-l),\quad &if \quad $\ \ l<x<d$,\cr
0,\quad &if \quad $-l<x<l$,\cr
\dot B_e(x+l),\quad &if \quad $-d<x<-l$,
}
\label{e38}
\end{equation}
where the magnetic field penetration depth, l, is equal to
\begin{equation}
l=d-{B_e\over\mu_0 j_c}.
\label{e39}
\end{equation}
\par
We consider now the stability of the stationary electric 
field and temperature distributions corresponding to the 
Bean's critical state against small electric field and 
temperature perturbations. To do it let us present the 
electric field $E(x,t)$ and the temperature $T(x,t)$ in the 
following form
\begin{equation}
E(x,t)=E_b(x)+\delta\! E(x)=
E_b(x)+\epsilon (x)\,\exp(\gamma t),
\label{e40} 
\end{equation}
\begin{equation}
T(x,t)=\tilde T_0+\delta T(x)=
\tilde T_0+\theta (x)\,\exp(\gamma t),
\label{e41}
\end{equation}
where 
\begin{equation}
\tilde T_0=T_0+{\dot B_eB_e^2\over 2h\mu_0^2j_c},
\label{e41a}
\end{equation} 
$\epsilon (x)\ll E_b(x)$, $\theta (x)\ll T_0$,
$({\rm Re}\,\gamma)^{-1}$ is the characteristic time of 
the magneto-thermal instability increase, and 
${\rm Im}\,\gamma$ is the frequency of the magneto-thermal 
oscillations. The stationary temperature $\tilde T_0$ is 
different from $T_0$ due to the Joule heating power 
$j_cE_b$ produced by the background electric field $E_b$. 
Let us note, that the difference between $\tilde T_0$ and 
$T_0$ is small, {\it i.e.}, $\tilde T_0-T_0\ll T_c-T_0$. 
Indeed, using Eq.~(\ref{e29}) we estimate the value of 
$\tilde T_0-T_0$ as
$\tilde T_0-T_0\approx (T_c-T_0)/n\ll T_c-T_0$.
\par
The small temperature 
$\delta T(x)=\theta(x)\,\exp(\gamma t)$ and electric 
field $\delta\! E=\epsilon (x)\,\exp(\gamma t)$
perturbations  decay if ${\rm Re}\,\gamma <0$. Therefore, 
the stability margin of the Bean's critical state is 
determined by the condition ${\rm Re}\,\gamma <0$.
\par
Substituting Eqs.~(\ref{e40}) and (\ref{e41}) into the 
Maxwell equation and the heat diffusion equation 
\begin{eqnarray}
\kappa\,{\partial^2 T\over\partial x^2}+j_cE&=&
C{\partial T\over\partial t},\\ 
\label{e42}
{\partial^2 E\over\partial x^2}&=&
\mu_0\,{\partial j\over\partial t}
\label{e43}
\end{eqnarray}
we find the system of equations describing $\epsilon (x)$
and $\theta (x)$. It takes the form
\begin{eqnarray}
\theta ''-{\gamma C\over\kappa}\,\theta
&=&-{j_c\over\kappa}\,\epsilon
\label{e44}\\
\epsilon''-{\mu_0\gamma j_c\over nE_b(x)}\,\epsilon
&=&-{\mu_0\gamma j_c\over T_c-T_0 }\,\theta,
\label{e45}
\end{eqnarray}
where the prime is to denote the derivative over x. Note, 
that to derive Eq.~(\ref{e45}) we use Eq.~(\ref{e38}) and 
the relation
\begin{equation}
{\partial j\over\partial t}
={\partial j\over\partial E}\,{\partial E\over\partial t}-
\Big\vert{\partial j_c\over\partial T}\Big\vert\,
{\partial T\over\partial t}=
{\gamma j_c\over nE_b}\,\epsilon-
{\gamma j_c\over T_c-T_0}\,\theta.
\label{e46}
\end{equation}
\par
We assume that the superconducting slab is in a thermal 
contact with a coolant with the temperature $T_0$ and we 
treat here the case when the external magnetic field ramp 
rate is given, {\it i.e.}, $E'(\pm d)=\dot B_e$. In 
addition, the electric field $E(x)$ is equal to zero in the 
inner region of the superconducting slab ($|x|\le l$). As a 
result, the boundary conditions for Eqs.~(\ref{e44}) and 
(\ref{e45}) are given by
\begin{equation}
\theta'(\pm d)=\mp\,{h\over\kappa}\,\theta(\pm d),
\label{e47}
\end{equation}
\begin{equation}
\epsilon(\pm l)=0,\quad\epsilon'(\pm d)=0.
\label{e48}
\end{equation}
\par
Let us present the solution for $\epsilon (x)$ in the form
\begin{equation}
\epsilon={n\theta\over T_c-T_0}\,E_b(x)+\epsilon_1(x),
\label{e49} 
\end{equation} 
where the first term corresponds to the approximation of a 
``frozen-in'' magnetic flux (see Eq.~(\ref{e13})) and the
second term describes the deviation from this 
approximation. It follows from Eqs.~(\ref{e45}), 
(\ref{e48}), and (\ref{e49}) that with the accuracy of 
$Bi\ll 1$ the equation for $\epsilon_1(x)$ has the form
\begin{equation}
\epsilon''_1-{\mu_0\gamma j_c\over n\dot B_e(|x|-l)}\,
\epsilon_1 =0
\label{e50}
\end{equation}
with the boundary conditions
\begin{equation}
\epsilon_1(\pm l)=0,\quad
\epsilon'_1(\pm d)=-{n\dot B_e\theta\over T_c-T_0}.
\label{e51} 
\end{equation}
\par
We consider here the case of $\tau\gg 1$, {\it i.e.}, the 
case when in the first approximation in $\tau^{-1}\ll 1$ 
the magnetic flux is ``frozen-in'' in the bulk of the 
superconducting slab. It means that during the rapid 
heating stage the magnetic flux redistribution takes place 
only in a thin surface layer with the thickness 
$\delta_s\ll d-l$. In other words, the function 
$\epsilon_1(x)$ decays inside the superconducting slab and
differs from zero only if $d-|x|$ is less or of the order 
of the skin depth $\delta_s$. In the region $d-|x|\ll d-l$ 
Eq.~(\ref{e50}) takes the form
\begin{equation}
\epsilon''_1-{\gamma B_p^2\over n\dot B_eB_ed^2}\,
\epsilon_1=0.
\label{e52} 
\end{equation} 
\par
The solution of Eq.~(\ref{e52}) matching the boundary 
conditions given by Eq.~(\ref{e51}) reads
\begin{equation}
\epsilon_1(x)=-{n\dot B_e\delta_s\theta\over T_c-T_0}\,
\exp \Bigl({|x|-d\over\delta_s}\Bigr),
\label{e53} 
\end{equation}
where we introduce the value of the skin depth $\delta_s$ 
as
\begin{equation}
\delta_s=d\sqrt{nB_e\dot B_e\over\gamma B_p^2}.
\label{e54} 
\end{equation}
\par
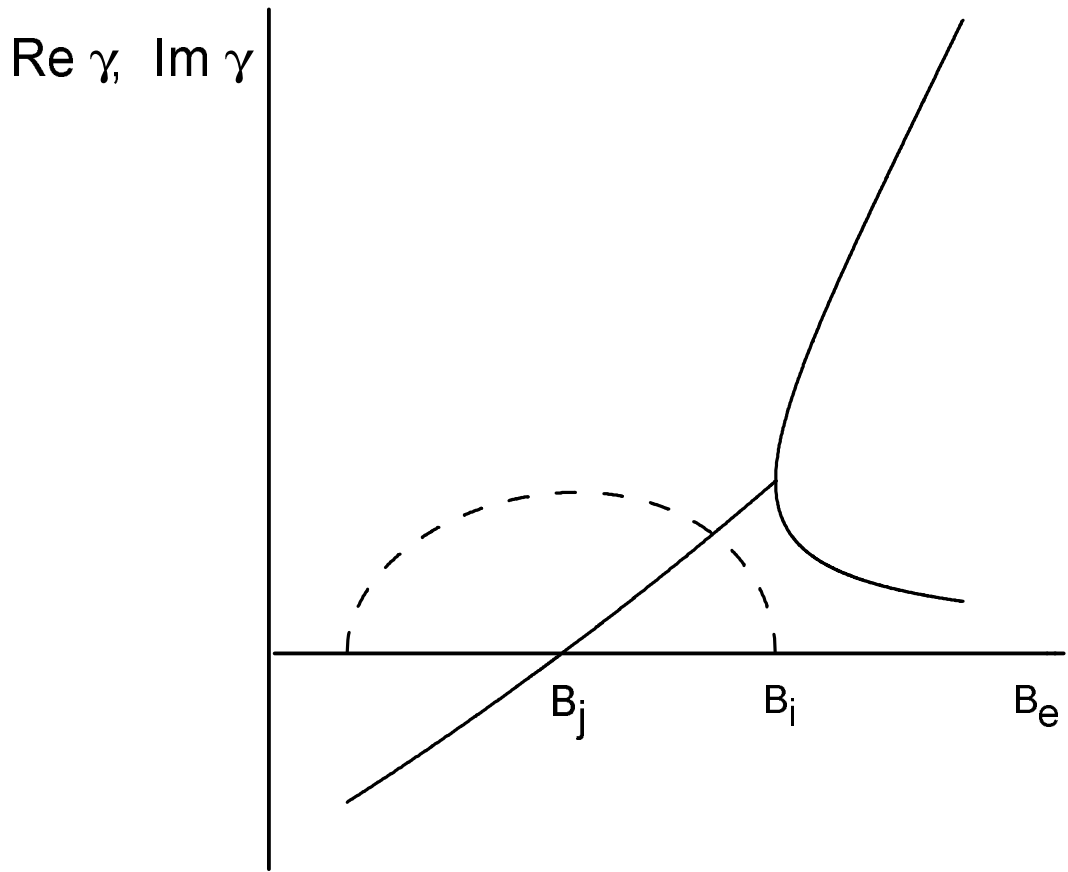
\begin{figure}
\caption{The dependencies of ${\rm Re}\,\gamma$ 
(solid line) and ${\rm Im}\,\gamma$ (dashed line) on 
$B_e$.}
\label{f2}
\end{figure}

To find the values of ${\rm Re}\,\gamma$ and 
${\rm Im}\,\gamma$ we integrate now Eq.~(\ref{e44}) over 
$x$ from $-d$ to $d$. Using Eqs.~(\ref{e47}), (\ref{e49}), 
and (\ref{e53}) we find the equation determining $\gamma$ 
in the form:
\begin{equation}
h-{n\dot B_eB_e^2\over 2\mu_0^2j_c(T_c-T_0)}=
-\gamma Cd-{B_en^2\dot B_e^2\over 
\gamma\mu_0^2j_c(T_c-T_0)}.
\label{e55} 
\end{equation}
We show schematically the dependencies of 
${\rm Re}\,\gamma$ and ${\rm Im}\,\gamma$ on $B_e$ in 
Fig.~\ref{f2}, where the field $B_i$ is determined by the 
equation
\begin{equation}
{B_i^2\over B_j^2}-1=
\sqrt{{8Cd\over h}\,{n\dot B_eB_i\over B_j^2}}.
\label{e55a} 
\end{equation}
The difference between $B_i$ and $B_j$ is small in the 
case when the ramp rate $\dot B_e$ is low, {\it i.e.},
$B_i-B_j\ll B_j$ if
\begin{equation}
\dot B_e<{Bi\over 2\pi^{2/3}n}\,{B_p\over t_{\kappa}}\,
\Bigl({B_a\over B_p}\Bigr)^{2/3}.
\label{e55b}
\end{equation} 
\par
It follows from Eq.~(\ref{e55}) that ${\rm Re}\,\gamma =0$
if $B_e=B_j$, {\it i.e.}, the Bean's critical state is 
stable if $B_e<B_j$, where the flux jump field $B_j$ is 
given by the Eq.~(\ref{e31}). We find also that at the 
stability threshold (for $B_e=B_j$) the value of $\gamma$ 
is imaginary, {\it i.e.}, $\gamma =i\omega$. Thus the 
magneto-thermal instability is preceded by the 
magneto-thermal oscillations with the frequency, 
$\omega$, given by the formula
\begin{equation}
\omega=\Biggl({2n^3\dot B_e^3h\over
\mu_0^2j_cd^2C^2(T_c-T_0)}\Biggr)^{1/4}
\propto \dot B_e^{\,3/4}
\label{e56} 
\end{equation}
\par
The ``frozen-in'' magnetic flux approximation is valid if
the surface layer where $\epsilon_1(x)\ne 0$ is thin, 
{\it i.e.}, if $\delta_s\ll d-l$. Using Eqs.~(\ref{e32}), 
(\ref{e39}), (\ref{e54}), and (\ref{e56}) we find the 
applicability criterion of the above approach in the form
\begin{equation}
B_j\gg B_a\,\Bigl({B_p\over\pi^2B_a}\Bigr)^{\,1/3}.
\label{e54a} 
\end{equation}
\par
\section{Summary}
To summarize, we consider the flux jump instability of 
the Bean's critical state in type-II superconductors. We 
show that under the conditions typical for most of the 
magnetization experiments this instability arise in the 
flux creep regime. We find the flux jump field $B_j$ 
that determines the critical state stability criterion. 
We show that the Bean's critical state stability is 
determined by the slope of the current-voltage curve. We 
calculate the dependence of $B_j$ on the external 
magnetic field ramp rate $\dot B_e$. We find the frequency 
of the magneto-thermal oscillations preceding a flux jump 
as a function on the external magnetic field ramp rate 
$\dot B_e$.
\par
\acknowledgments{
I am grateful to I.~Rosenman and L.~Legrand for useful 
discussions stimulated this work.}
\par

\end{document}